\documentclass[11pt]{article}
\usepackage{epsfig}
\newcommand{\arcmin}{^{\prime}}
\title{{\bf The Influence of the Starburst\\ on the ISM in NGC~1569}}
\author{S.~M\"uhle$^1$, S.~H\"uttemeister$^{1,2}$, U.~Klein$^1$, E.M.~Wilcots$^3$\\
\vspace{0.1cm}\\
\normalsize $^1$Radioastronomisches Institut Univ. Bonn, Auf dem H\"ugel 71, D-53121 Bonn, Germany\\
\normalsize $^2$Astronomisches Institut der Ruhr-Universit\"at, Postfach 102148, D-44780 Bochum, Germany\\
\normalsize $^3$Department of Astronomy, University of Wisconsin-Madison, Madison, WI 53706, USA\\
}
\date{}
\begin{document}
\maketitle
\def\bull{\vrule height .9ex width .8ex depth -.1ex}
\makeatletter
\def\ps@plain{\let\@mkboth\gobbletwo
\def\@oddhead{}\def\@oddfoot{\hfil\tiny
``Dwarf Galaxies and their Environment'';
Bad Honnef, Germany, 23-27 January 2001; Eds.{} K.S. de Boer, R.-J.Dettmar, U. Klein; Shaker Verlag}%
\def\@evenhead{}\let\@evenfoot\@oddfoot}
\makeatother

%%  if your contribution is short, you may, if the title is clear enough, 
%%  skip the abstract.....
%\begin{abstract}\noindent
%How Dwarf Galaxies got to Bad Honnef and how observational data and 
%the analyses came to be debated by many astronomers from several countries
%\end{abstract}

%%%%%%%%%%%%%%%%%%%%%%%%%%%%%%%%%%%%%%%%%%%%%%%%%%%%%%%%%%%%%%%%%%%%%%%%%%%%%%%

\section{Introduction}

The gas-rich dwarf irregular galaxy NGC~1569 (UGC 3056, Arp 210, VII Zw 16) is 
a member of the IC~342/Maffei galaxy group (Karachentsev et al.\ 1994) and lies 
near the galactic plane at a distance of about 
$(2.2 \pm 0.6)\,{\rm Mpc}$ (Israel 1988). Its closest neighbor is UGCA 92 at a 
projected distance of about 40 kpc ($\sim 1.2^{\circ}$). As can be expected 
of a 
dwarf galaxy, NGC~1569 is metal-poor ($Z=0.23\,{\rm Z_{\odot}}$, Gonz\'alez Delgado et 
al.\ 1997). There is much evidence that this galaxy currently is in 
a post-starburst phase (see next chapter). Therefore, it has been chosen as 
the target of a comprehensive study of the influence of a starburst on the ISM 
and the magnetic structure in a dwarf galaxy, which is the topic of 
S.~M\"uhle's Ph.D.\ thesis.
%that is performed in the context of the Graduiertenkolleg. 
Here we present our first results on the impact of the 
recent starburst on the neutral atomic and molecular gas.

%%%%%%%%%%%%%%%%%%%%%%%%%%%%%%%%%%%%%%%%%%%%%%%%%%%%%%%%%%%%%%%%%%%%%%%%%%%%%%%

\section{Evidence for a Starburst}

Among the most prominent features in NGC~1569 are two very luminous super-star 
clusters (e.g.\ O'Connell et al.\ 1994) that coincide with the centers of the 
distribution of the young stellar population. The star formation 
history as derived from photometric studies (e.g. Aloisi et al.\ 2001, Greggio 
et al.\ 1998, Vallenari \& Bomans 1996) indicates a starburst, a 
phase with a very high star formation rate of 
$0.5 \ldots 3 {\rm M_{\odot}/yr}$ 
($4 \ldots 20 {\rm M_{\odot}/yr\,kpc^2}$), from about 100 million years until 
about 4 million years ago, with no quiescent periods that lasted longer than 
about 
10 million years. Before the starburst, the overall star formation rate must 
have been quite low, as the large HI reservoir and the low metallicity of 
NGC~1569 indicate.

Independent evidence for a recent starburst has been found in the rather 
strong synchrotron emission of NGC~1569. With a delay of about $10^7$ years, 
a starburst leads to an increased rate of supernova explosions, each of them 
injecting a large amount of relativistic electrons into the ISM. These 
high-energy particles rapidly lose energy due to radiation losses. At the end 
of a starburst, when the supernova rate decreases, 
%when the reservoir of relativistic electrons isn't replenished by the same big number of supernovae any more, 
the synchrotron spectrum ages, 
visible as a distinct kink in the slope of the spectrum. In NGC~1569, Israel 
\& de Bruyn (1988) have found such a kink in the non-thermal continuum 
emission at 8 GHz, corresponding to 
%a time estimate for the 
an end of the starburst about 5 million years before now.

The ionized gas also reveals signs of strong recent star formation: expanding 
H$\alpha$ shells (Tomita et al.\ 1994) 
%and two arc-like structures (Waller 1991) 
have been detected in the disk of NGC~1569, and the H$\alpha$ image by 
Hunter et al.\ (1993) shows numerous filaments along the minor axis extending 
into the halo. In ROSAT observations, extended soft X-ray emission ($T\sim 
10^7\,{\rm K}$) has been detected at the location of the super-star clusters 
(Heckman et al.\ 1995, Della Ceca et al.\ 1996). Comparing the pressure of this
hot gas with the shallow gravitational potential, it becomes 
clear that this gas will most likely be blown away from the galaxy (Martin 1999).

%%%%%%%%%%%%%%%%%%%%%%%%%%%%%%%%%%%%%%%%%%%%%%%%%%%%%%%%%%%%%%%%%%%%%%%%%%%%%%%
%
%                     Figures 1 and 2

\begin{figure}[tb]
  \parbox{16.5cm}{
 \parbox{8.0cm}{
 \epsfig{file=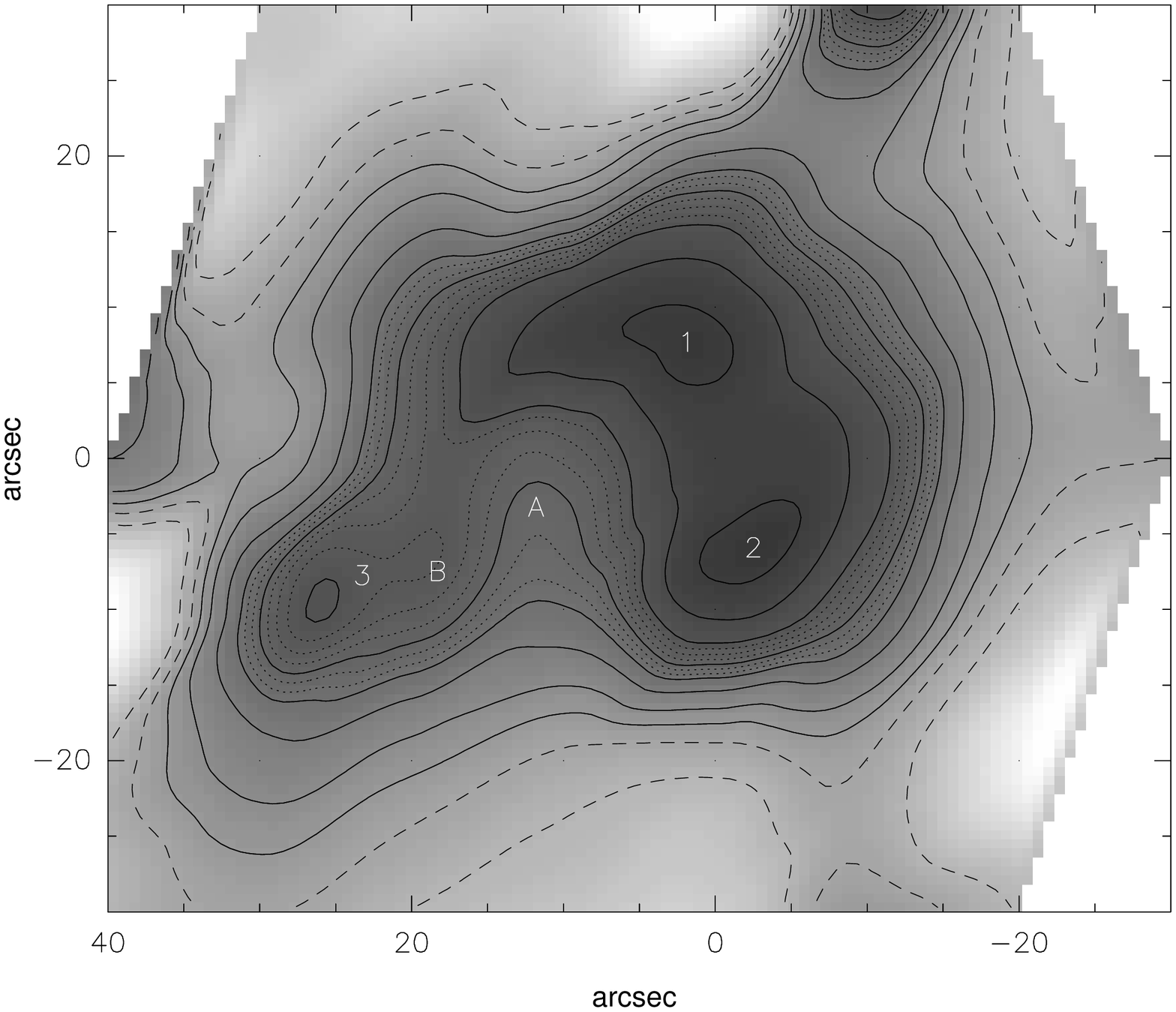,height=8cm,angle=-90}
 \caption{\small CO(3--2) map of the starburst region in NGC~1569. The contour
levels are 0.3, 0.7 (dashed), 1.0 (3$\sigma$), 1.3, 1.7, 2.0, 2.3, 2.7, 3.0 and $3.3\,{\rm K\,km/s}$. The dotted lines trace 2.1, 2.2, 2.4, 2.5 and $2.6\,{\rm K\,km/s}$. The positions of the super-star clusters are denoted ``A'' and ``B''.}}
 \parbox{0.3cm}{\hspace*{0.3cm}}
 \parbox{8.0cm}{
 \epsfig{file=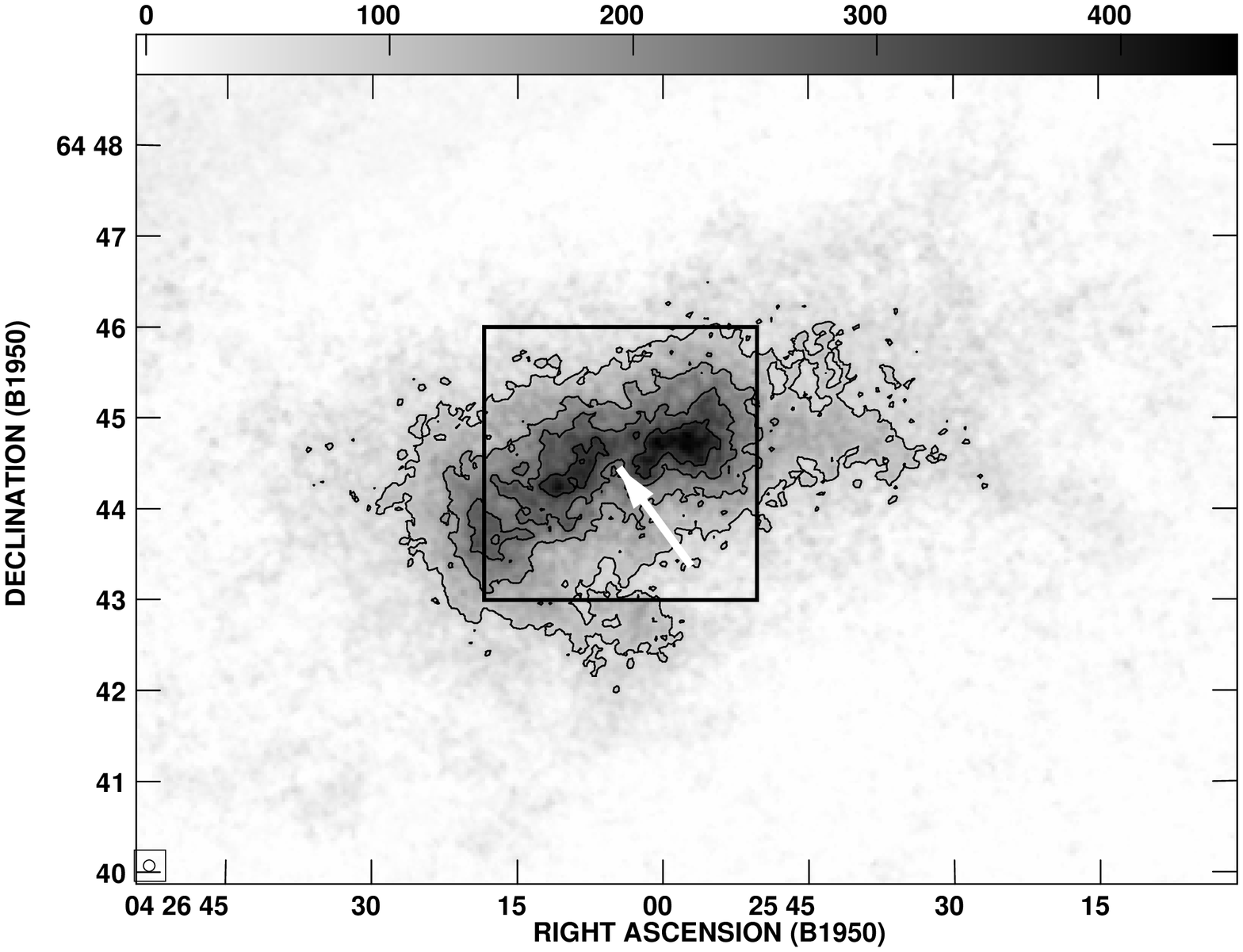,height=8cm, angle=-90}
 \caption[]{\small High-resolution HI column density map. The contour lines are 20$\sigma$, 40$\sigma$, 60$\sigma$ and 80$\sigma$. The peak value of $447\,{\rm Jy/bm\;m/s}$ corresponds to 103$\sigma$. The white arrow indicates the location of the super-star clusters, the black square an area of $3\arcmin\times3\arcmin$ (about $2\,{\rm kpc}\times2\,{\rm kpc}$).}}}
\end{figure}

%%%%%%%%%%%%%%%%%%%%%%%%%%%%%%%%%%%%%%%%%%%%%%%%%%%%%%%%%%%%%%%%%%%%%%%%%%%%%%%

\section{The Impact on the Molecular Gas}

In the course of this study, we have mapped the CO(3--2) emission of NGC~1569 
near the 
super-star clusters with the Heinrich-Hertz Telescope (Figure 1). The map 
shows not only three giant molecular associations (GMAs), but also much 
extended emission. The depression in the CO emission at the position of the 
super-star cluster A hints at a hole in the distribution of the molecular gas, 
but might as well be caused by dissociation of the CO molecules in the strong 
radiation field near the super-star cluster. The GMA \#2 in our map coincides 
nicely with the giant molecular clouds (GMCs) \#1 and 2 found by Taylor et al.\ 
(1999) in maps of the 
CO(2--1) and CO(1--0) transitions obtained with the Plateau de Bure 
interferometer. Our northern GMA (\#1) can be associated with the GMCs \#4 
and 5. 
%in the CO(1--0) map. 
The GMAs \#1 and 2 also coincide with the rim of a
high-density HI ridge. A comparison of the fluxes of the interferometric data 
with the few available single-dish spectra (Greve et 
al.\ 1996) shows that a substantial fraction of the total emission 
%must origin from diffuse gas and thus be 
is missing in these maps (Taylor et 
al.\ 1999). A first analysis of the single-dish data at a position
between the GMAs \#1 and 2 using our CO(3--2) map as well as the data by 
Greve et al.\ (1996) results in rather high line ratios
$$ \frac{I_{\rm CO(2-1)}}{I_{\rm CO(1-0)}} = 1.1 \quad {\rm (Greve\ et\ al.\ 
1996)\ and} \quad\quad \frac{I_{\rm CO(3-2)}}{I_{\rm CO(2-1)}} = 1.4,$$
indicating an unusually warm molecular gas phase. An in-depth non-LTE analysis
of the physical properties of the giant molecular clouds has to wait for new 
single-dish observations of the CO
%CO(2--1), CO(1--0), $^{13}$CO(2--1) and $^{13}$CO(2--1) 
emission to be performed in March and April 2001.

%%%%%%%%%%%%%%%%%%%%%%%%%%%%%%%%%%%%%%%%%%%%%%%%%%%%%%%%%%%%%%%%%%%%%%%%%%%%%%%
%
%                     Figures 3 and 4

\begin{figure}[tb]
  \parbox{16.5cm}{
 \parbox{8.0cm}{
 \epsfig{file=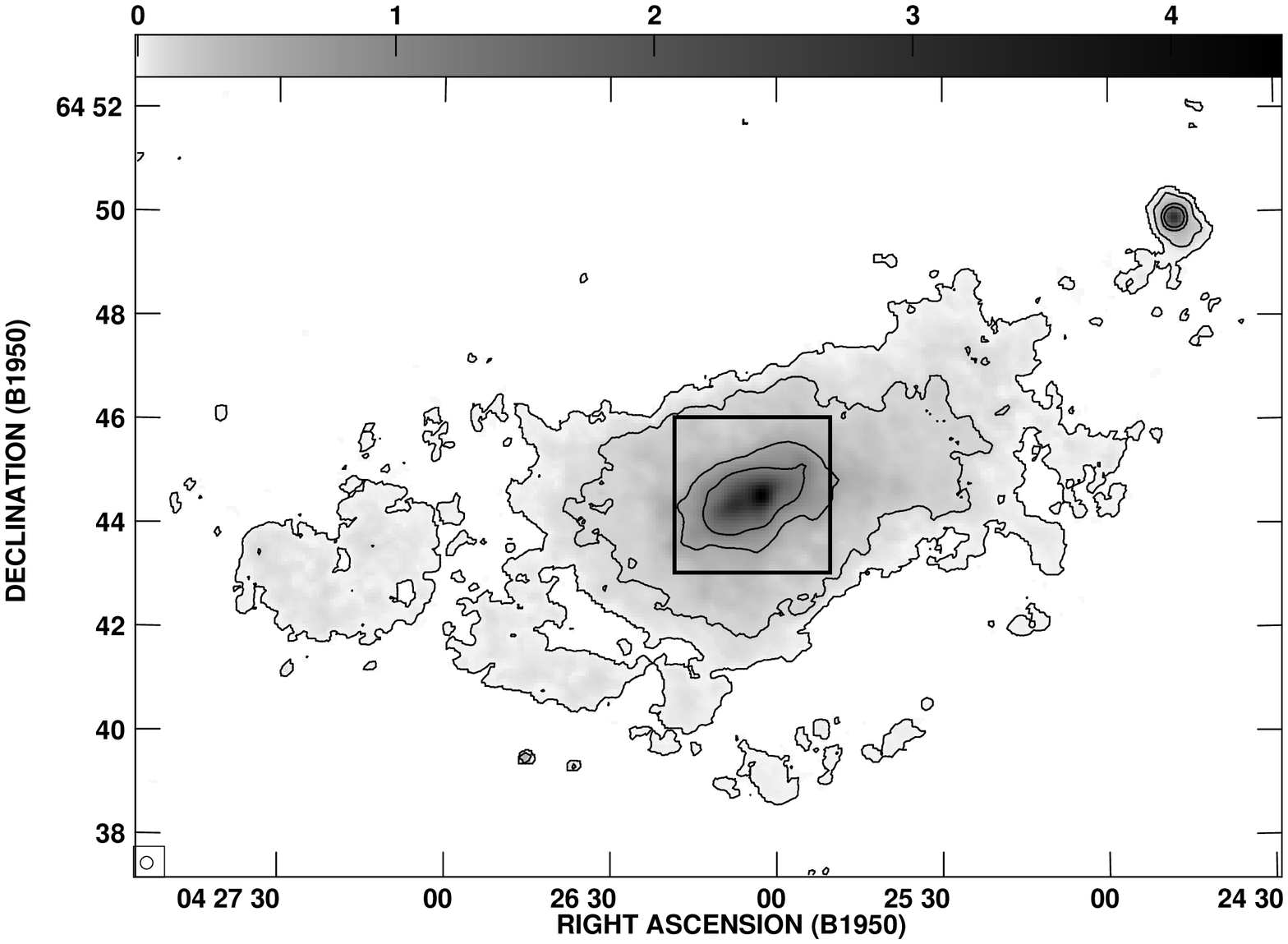,height=8cm,angle=-90}
 \caption{\small Low-resolution HI column density map. The contour lines are 5$\sigma$, 100$\sigma$, 500$\sigma$ and 1000$\sigma$. The peak value of $4421\,{\rm Jy/bm\;m/s}$ corresponds to 4000$\sigma$. The square has an extent of $3\arcmin\times3\arcmin$ (about $2\,{\rm kpc}\times2\,{\rm kpc}$). }}
 \parbox{0.3cm}{\hspace*{0.3cm}}
 \parbox{8.0cm}{
 \epsfig{file=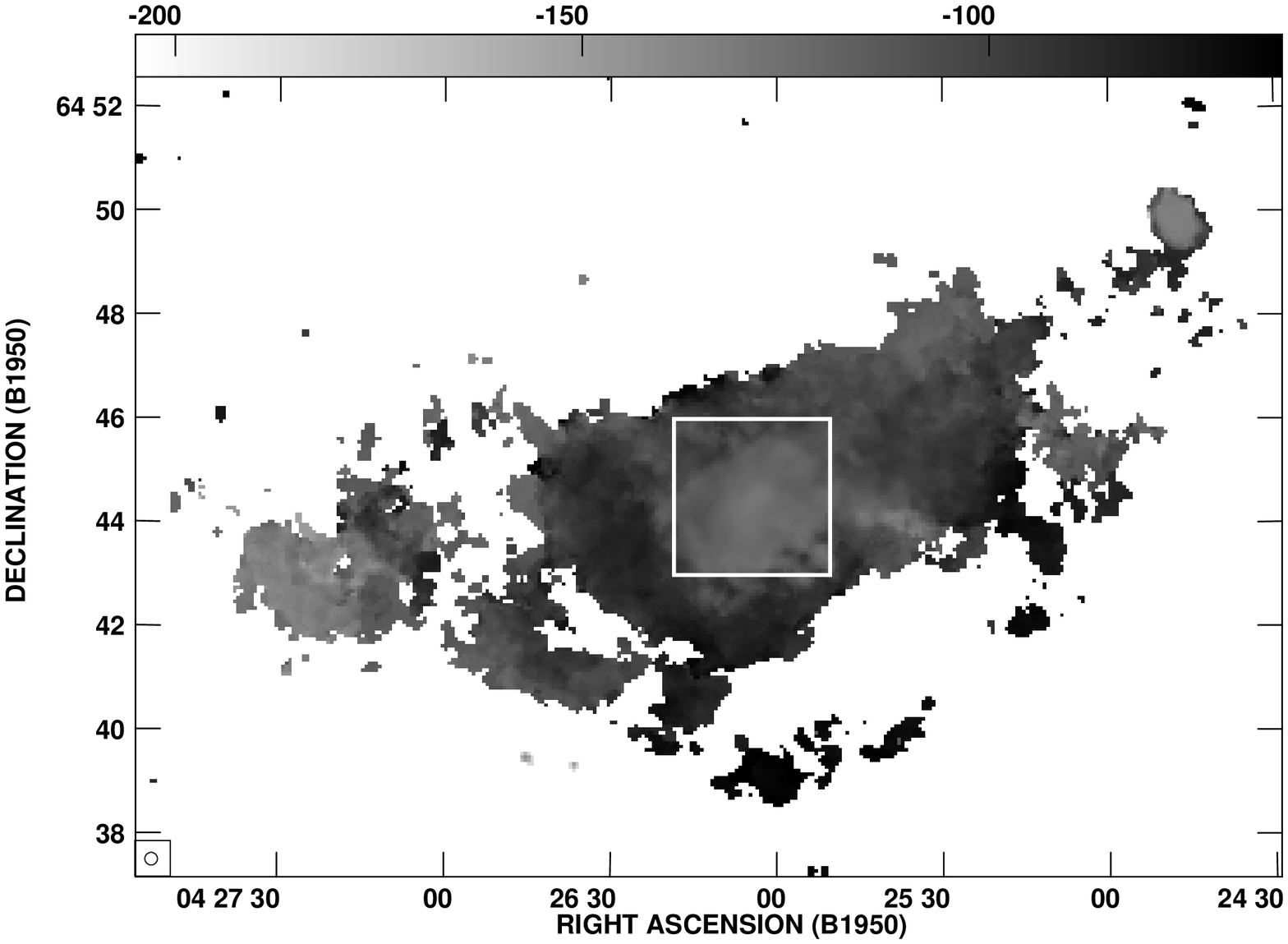,height=8cm, angle=-90}
 \caption[]{\small The velocity field of the low-resolution HI cube. The radial velocities range from $-205\,{\rm km/s}$ to $-64\,{\rm km/s}$. Higher velocities have been truncated to avoid Galactic emission. The square has an extent of $3\arcmin\times3\arcmin$ (about $2\,{\rm kpc}\times2\,{\rm kpc}$).}}}
%\epsfig{file=yourfig1.ps,width=12cm}
%\caption[]{Your caption} 
\end{figure}

%%%%%%%%%%%%%%%%%%%%%%%%%%%%%%%%%%%%%%%%%%%%%%%%%%%%%%%%%%%%%%%%%%%%%%%%%%%%%%%

\section{The Impact on the Atomic Gas}

Dwarf galaxies with violent star formation usually have a complex HI 
distribution with a lot of holes and arcs (e.g.\ Walter \& Brinks 1999). 
Surprisingly, the column density map of the high-resolution HI cube (Figure 2), 
that we have obtained from observations with the VLA B-, C- and D-arrays, 
shows a remarkably smooth distribution of the neutral gas, with only one 
prominent hole and two arm-like extensions in the west and in the south of the 
disk. Searching carefully for holes in the channel maps and position-velocity 
plots of this cube, we have found only two holes: The first hole is the one 
already detected in the HI column density map. Adopting the classification 
scheme of Brinks \& Bajaja (1986), this is a type 1 hole with a diameter of 
about $1\,{\rm kpc}$ and an expansion velocity of about $7\,{\rm km/s}$. 
It is associated with the two super-star clusters and an estimate of its 
kinematic age assuming a 
constant expansion velocity yields $t=8\cdot10^7\,{\rm yr}$, supporting a 
causal connection between the starburst and the creation of this hole. The 
channel maps also show that the center of the hole is offset to the south of 
the position of the super-star clusters, suggesting different expansion velocities in the 
high-density HI ridge and in the thinner gas in the southern part of the disk.
The other hole is classified as a type 2 hole 
%because the low-velocity rim of the hole is missing in the position-velocity 
%diagram, 
and is located southeast 
of the first hole. With a diameter of about $600\,{\rm pc}$ and an estimated 
expansion velocity of about $28\,{\rm km/s}$, it is kinematically younger 
($t=1\cdot10^7\,{\rm yr}$). We label this hole a candidate hole, since 
it is located outside the optical extent of the disk and might be a chance 
feature formed by the rim of the dense HI distribution and the low-mass HI arm 
described in the next paragraph.

Trading the high-resolution of our cube for increased sensitivity in our 
low-resolution HI column density map (Figure 3), we 
have detected a low-mass HI cloud (``companion'') east of the HI distribution 
of NGC~1569 as well as an HI bridge connecting the cloud with the main HI 
disk. The measured column densities translate to mass estimates of 
$2.4\cdot10^7\,{\rm M_{\odot}}$ for the companion and 
$2.4\cdot10^8\,{\rm M_{\odot}}$ for the disk of NGC~1569. This is in good 
agreement with the findings of Stil \& Israel (1998). As to the origin of the 
companion, that has no known optical counterpart (Stil \& Israel 1998), and 
the associated bridge, we have investigated different possibilities: In the 
velocity field (Figure 4), we see a 
clear offset of the radial velocities of arm and companion from the velocity 
structure of the disk. Thus, it is very unlikely that these features form an 
outer part of the main HI distribution. The nearest neighbor to NGC~1569, 
UGCA~92 is a dwarf galaxy at a similar radial velocity, but less massive than 
NGC~1569. Comparing the gravitational 
force that NGC~1569 exerts on the companion with that of the far-away 
neighbor, we can discard the scenario that the companion and the bridge are 
tidal features pulled out of the disk of NGC~1569. Another hypothesis is that 
these features are part of an HI cloud falling into the disk of 
NGC~1569 and being tidally disrupted in this process. Since the potential 
energy of the companion is of the same order of magnitude as its kinetic 
energy, 
we consider this a possible explanation for the origin of the HI features
southeast of the main disk of NGC~1569 and worth further investigation.

The sensitive low-resolution column density map and velocity field also reveal 
other interesting details: Apart from a radio continuum source northwest of 
NGC~1569, especially the velocity field clearly shows the remnants of a huge 
HI shell reaching from the center of the disk out into the halo in the 
southwest. The western arm detected in the high-resolution column density map
seems to be part of this huge shell. The southern feature found in the same 
map may be interpreted as the continuation of the HI bridge.

%%%%%%%%%%%%%%%%%%%%%%%%%%%%%%%%%%%%%%%%%%%%%%%%%%%%%%%%%%%%%%%%%%%%%%%%%%%%%%%

\section{Summary \& Work in Progress}

In the nearby dwarf galaxy NGC~1569, evidence for a recent strong 
starburst has been found in the star formation history as well as in the 
synchrotron spectrum and in the distribution of the ionized gas. We propose 
that the starburst has had a strong impact on the neutral atomic and molecular 
gas, too: The distribution of both the HI and the CO emission show a 
depression at the location of the two super-star clusters. The surrounding 
molecular gas seems to be unusually warm. In the HI column density map, we 
have detected an extended feature that might be an infalling HI cloud, a 
possible trigger for the starburst.\\
A next step will be the combination of our VLA 
data with single-dish HI observations in order to get the complete picture of 
the distribution of neutral hydrogen in NGC~1569. For a thorough non-LTE 
investigation 
of the physical properties of the giant molecular clouds near the super-star 
clusters, we will map the starburst region in the CO(2--1) and CO(1--0) 
transitions with a single-dish telescope and observe the $^{13}$CO(2--1) and 
$^{13}$CO(1--0) emission of selected areas. Another project that has just been 
started in the context of this Ph.D. work is the determination of the 
polarization, the magnetic structure and field strength in the halo of 
NGC~1569. This will allow us to include the effect of a magnetic field in
the feedback processes of a starburst.

%%%%%%%%%%%%%%%%%%%%%%%%%%%%%%%%%%%%%%%%%%%%%%%%%%%%%%%%%%%%%%%%%%%%%%%%%%%%%%%

% References:
%%  Use the ApJ, AJ, new A\&A style (are the same!)
%%
{\small
\begin{description}{} \itemsep=0pt \parsep=0pt \parskip=0pt \labelsep=0pt
\item {\bf References}

\item 
Aloisi, A., Clampin, M., Diolaiti, E., et al. 2001, AJ in press (March 2001)
\item 
Brinks, E., \& Bajaja, E. 1986, A\&A 169, 14
\item 
Della Ceca, R., Griffiths, R.E., Heckman, T.M., \& Mac~Kenty, J.W. 1996, ApJ 469, 662
\item
Gonz\'alez Delgado, R.M., Leitherer, C., Heckman, T., \& Cervi\~no, M. 1997, ApJ 483, 705
\item 
Greggio, L., Tosi, M., Clampin, M., et al. 1998, ApJ 504, 725
\item
Greve, A., Becker, R., Johansson, L.E.B., \& Mc~Keith, C.D. 1996, A\&A 312, 391
\item 
Heckman, T.M., Dahlem, M., Lehnert, M.D., Fabbiano, G., Gilmore, D., \& Waller, W.H. 1995, ApJ 448, 98
\item 
Hunter, D.A., Hawley, W.N., \& Gallagher, J.S. 1993, AJ 106, 1797 
\item 
Israel, F.P. 1988, A\&A 194, 24
\item 
Israel, F.P., \& de Bruyn, A.G. 1988, A\&A 198, 109
\item 
Karachentsev, I.D., Tikhonov, N.A., \& Sazonova, L.N. 1994, AstL 20, 84  
\item 
Martin, C.L. 1999, ApJ 513, 156
\item
O'Connell, R.W., Gallagher, J.S., \& Hunter, D.A. 1994, ApJ 433, 65
%\item
%Reakes, M. 1980, MNRAS 192, 297
\item
Stil, J.M., \& Israel, F.P. 1998, A\&A 337, 64
\item 
Taylor, C.L., H\"uttemeister, S., Klein, U., \& Greve, A. 1999, A\&A 349, 424
\item 
Tomita, A., Ohta, K., \& Saito, M. 1994, PASJ 46, 335
\item 
Vallenari, A., \& Bomans, D.J. 1996, A\&A 313, 713
%\item 
%Waller, W.H. 1991, ApJ 370, 144
\item 
Walter, F., \& Brinks, E. 1999, AJ 118, 273 

%\item
%Klein, U., Dettmar, R.-J., \& de Boer, K.S. 2001, in 
%``Dwarf Galaxies and their Environment'', eds K.S.~de~Boer, R.J.~Dettmar, 
%U.~Klein, Shaker Verlag; in press
%\item
%Next, X. 1999, A\&A 345, 1234

\end{description}
}

\end{document}